\documentclass[aps,pre,eqsecnum,preprint,tightenlines,nofootinbib,showpacs]{revtex4-1}
\usepackage{graphicx,amsmath,latexsym}

\def\bea{\begin{eqnarray}}
\def\eea{\end{eqnarray}}
\def\be{\begin{equation}}
\def\ee{\end{equation}}
\newcommand{\ub}[1]{\underline{#1}}

\newcommand{\Pminus}{{\cal P}^-}
\newcommand{\nsvd}{n_{\rm svd}}

\begin{document}

\title{Compression algorithm for discrete light-cone quantization
}

\author{Xiao Pu}
\author{Sophia S. Chabysheva}
\author{John R. Hiller}
\affiliation{Department of Physics \\
University of Minnesota-Duluth \\
Duluth, Minnesota 55812}

\date{\today}

\begin{abstract}

We adapt the compression algorithm of Weinstein, Auerbach, and Chandra
from eigenvectors of spin lattice Hamiltonians to eigenvectors of
light-front field-theoretic Hamiltonians.  The latter are approximated
by the standard discrete light-cone quantization technique, which
provides a matrix representation of the Hamiltonian eigenvalue
problem.  The eigenvectors are represented as singular value
decompositions of two-dimensional arrays, indexed by transverse
and longitudinal momenta, and compressed by truncation of the
decomposition.  The Hamiltonian is represented by a rank-four
tensor that is decomposed as a sum of contributions factorized
into direct products of separate matrices for transverse and
longitudinal interactions.  The algorithm is applied to a model
theory, to illustrate its use.

\end{abstract}

%

\maketitle

\section{Introduction}
\label{sec:introduction}

The nonperturbative solution of quantum field theories generally requires
numerical techniques.  In Hamiltonian formulations, these typically
take the form of matrix approximations to the eigenvalue problem for
the mass eigenstates.  The eigenstates themselves are represented by
one-dimensional vectors with multiple indices, such as momentum components.
If the Hamiltonian matrix elements are computed as needed and
not stored in computer memory, the dominant memory requirement
is the storage of these vectors.  To possibly reduce this memory
requirement, we wish to explore compression of these vectors.

For the case of Hamiltonians for spin lattices, Weinstein {\em et al}.~\cite{Weinstein}
have proposed a compression algorithm based on singular value decomposition
(SVD)~\cite{NumRecipes}.
The lattice is split into two, and the eigenvectors $|\psi\rangle$
become two-dimensional matrices indexed by the sublattices.  A singular 
value decomposition reduces the matrix to a sum over contributions from 
individual sublattice vectors
\be
|\psi\rangle=\sum_\alpha \lambda_\alpha|\alpha\rangle_1|\alpha\rangle_2,
\ee
where $|\alpha\rangle_i$ is a vector on the ith sublattice and $\lambda_\alpha$
is the associated singular value of the original matrix.
The sum over $\alpha$ is truncated to keep the $\nsvd$ largest contributions,
thereby storing a compressed matrix as $2\nsvd$ sublattice vectors.  By doing
all calculations in the compressed SVD format, memory requirements
are reduced.  There is, of course, a computational cost due to the extra
processing.  For diagonalization with a Lanczos-type iteration~\cite{Lanczos},
the key step is multiplication of a compressed vector by the Hamiltonian,
which must be done entirely within the SVD representation.  The Weinstein
algorithm includes this step.

A standard technique for the nonperturbative solution of a quantum
field theory is discrete light-cone quantization (DLCQ)~\cite{PauliBrodsky,DLCQreview}. 
The theory is quantized in light-cone coordinates~\cite{Dirac},
which we take to be the light-cone time coordinate $x^+=t+z$ and light-cone spatial
coordinates $\ub{x}=(x^-\equiv t-z,\vec{x}_\perp)$, with $\vec{x}_\perp=(x,y)$
being the transverse piece.  Quantization in terms of these coordinates
has the advantage of providing well-defined wave functions for eigenstates
of the Hamiltonian~\cite{DLCQreview}.  The wave functions appear
as coefficients in a Fock-state expansion of the eigenstate and are functions
of the light-cone momenta $\ub{p}_i=(p_i^+\equiv E_i+p_{iz},\vec{p}_{i\perp})$.
The eigenstate $|\psi(\ub{P})\rangle$ has total momentum $\ub{P}$ and satisfies
the field-theoretic Schr\"odinger equation
\be
\Pminus|\psi(\ub{P})\rangle=\frac{M^2+P_\perp^2}{P^+}|\psi(\ub{P})\rangle,
\ee
where $\Pminus$ is the light-cone Hamiltonian, $M$ is the mass of the state,
and the expression given as the eigenvalue represents the mass-shell condition
$P^-\equiv E+P_z=(M^2 +P_\perp^2)/P^+$.  The wave functions then satisfy integral
equations obtained from this fundamental Schr\"odinger equation.
The DLCQ approximation is roughly equivalent to a trapezoidal
approximation to the integrals; the wave functions are represented by
values at discrete momentum values and the Hamiltonian by a matrix.

The vector that represents wave-function values is indexed by longitudinal
and transverse momentum coordinates as well as other possible indices,
such as flavor and spin.  Therefore, it is easily re-interpreted as
a multidimensional structure with many indices.  By collecting these
indices into two disjoint sets, instead of the one original set,
the multi-ranked matrix becomes a rank-two matrix to which SVD can be applied.
Here we propose a separation between transverse and longitudinal momenta.
For the sample application that we use as illustration, this
is all that is needed.  More generally, one can imagine including
spin indices with transverse momenta and flavor indices with longitudinal
momenta, to facilitate the factorization of the Hamiltonian.
Angular momentum conservation in the $z$ direction naturally 
pairs light-cone spin with transverse momentum in interactions, and
flavor-dependent interactions in light-cone Hamiltonians commonly 
involve the constituent mass and longitudinal momenta. 

The SVD decomposition creates two sets of vectors, one with transverse
index and one with longitudinal index, and
the Hamiltonian matrix becomes a rank-four tensor.  However, the Hamiltonian
can be factorized as the sum of direct products of rank-two matrices,
with one matrix indexed by transverse momenta and the other by
longitudinal momenta.  Thus, the Hamiltonian acts separately on
the transverse vectors and longitudinal vectors.  Because the factorization
of the Hamiltonian typically has many terms, the outcome of
multiplication by the Hamiltonian must be summed over many more contributions
than in the original SVD decomposition and, therefore, must be compressed.
This is where there is a significant computational load.  However, this
is the price to pay for reductions in memory requirements.

In Sec.~\ref{sec:algorithm} we present the compression and diagonalization
algorithms in some detail, both for completeness and to establish some
notation. A sample application to a model theory is discussed in 
Sec.~\ref{sec:application}.  The results are summarized in the final
section, Sec.~\ref{sec:summary}. 

\section{Algorithm}
\label{sec:algorithm}

\subsection{SVD compression}
\label{sec:compression}

We present the Weinstein compression algorithm~\cite{Weinstein} with
expanded notation and in the context of the
field-theoretic applications that we have in mind.  The state vectors
are Fock-state expansions, and each Fock state is factorized into
a direct product of transverse and longitudinal states
\be
|\psi\rangle=\sum_n\sum_{i_nj_n}A_{i_nj_n}|v_{i_n}\rangle|w_{j_n}\rangle.
\ee
The coefficients $A_{i_nj_n}$ are the discrete amplitude values at
transverse momentum index $i_n$ and longitudinal momentum index $j_n$
in the $n$th Fock sector.
The $i_n$-$j_n$ Fock state has been factorized as $|v_{i_n}\rangle|w_{j_n}\rangle$.
In what follows, we combine all Fock sectors into the range of the indices,
rather than keeping a separate sum over sectors, and drop the subindex $n$;
however, in practice, the amplitudes $A_{ij}$ are compressed sector by sector,
because the Hamiltonian acts only between neighboring sectors.

The SVD decomposition of $A$ can be written as~\cite{NumRecipes}
\be
A_{ij}=\sum_{k=1}^{N_A} \lambda_k V_{ik} W_{jk}.
\ee
The singular values $\lambda_k$ are positive, decreasing in magnitude,
and normalized to one: $\sum_k\lambda_k^2=1$.  The columns of $V$ and $W$
are orthonormal:
\be
\sum_i V_{ik}^*V_{ik'}=\delta_{kk'},\;\;
\sum_j W_{jk}^*W_{jk'}=\delta_{kk'}.
\ee
The storage of the amplitudes is compressed by truncating 
the sum over $k$ at $\nsvd<N_A$, keeping only terms with
larger values of $\lambda_k$.

The generic Hamiltonian $H$ is factorized as a direct sum
\be
H=\sum_{\alpha=1}^{N_H} H_V^{(\alpha)}\otimes H_W^{(\alpha)}.
\ee
Its matrix elements in the factorized Fock basis are
\be
H_{ij,i'j'}=\sum_{\alpha=1}^{N_H} H_{Vii'}^{(\alpha)} H_{Wjj'}^{(\alpha)},
\ee
with
\be
H_{Vii'}^{(\alpha)}\equiv\langle v_i|H_V^{(\alpha)}|v_{i'}\rangle,\;\;
H_{Wjj'}^{(\alpha)}\equiv\langle w_j|H_W^{(\alpha)}|w_{j'}\rangle.
\ee
Thus, although the matrix elements of $H$ form a rank-four tensor, the
tensor is decomposed into a set of rank-two matrices.  The action of $H$ on
a state vector is to produce a new set of amplitudes
\be \label{eq:newA}
\tilde{A}_{ij}=\sum_{k=1}^{\nsvd}\sum_{\alpha=1}^{N_H}v_i^{(\alpha,k)}w_j^{(\alpha,k)},
\ee
with
\be
v_i^{(\alpha,k)}\equiv\sqrt{\lambda_k}\sum_{i'}H_{Vii'}^{(\alpha)}V_{i'k},\;\;
w_j^{(\alpha,k)}\equiv\sqrt{\lambda_k}\sum_{j'}H_{Wjj'}^{(\alpha)}W_{j'k}.
\ee
However, the new amplitude values are constructed from a sum over
$\nsvd N_H$ pairs of nonorthogonal vectors.  The new vectors must
be re-orthogonalized, and only the $\nsvd$ most important of them kept.
This is the key step in the Weinstein algorithm~\cite{Weinstein}, which
we now describe in our notation.

The new vectors are orthogonalized by application of SVD to the
overlap matrices
\be
\langle v^{(\alpha',k')}|v^{(\alpha,k)}\rangle=(U_V^\dagger D_V U_V)_{\alpha k,\alpha' k'}
\ee
and 
\be
\langle w^{(\alpha',k')}|w^{(\alpha,k)}\rangle=(U_W^\dagger D_W U_W)_{\alpha k,\alpha' k'},
\ee
where $U_V$ and $U_W$ are unitary and $D_V$ and $D_W$ are diagonal.  Notice also
that the indices on the right-hand sides are reversed in order.  Orthonormal
vectors can then be formed as
\be
\tilde v_i^{(\alpha',k')}=\sum_{\alpha,k}(D_V^{-1/2}U_V)_{\alpha'k',\alpha k}v_i^{(\alpha,k)}
\ee
and
\be
\tilde w_j^{(\alpha',k')}=\sum_{\alpha,k}(D_W^{-1/2}U_W)_{\alpha'k',\alpha k}w_j^{(\alpha,k)}.
\ee
Inversion of these sums recovers the original vectors
\be
v_i^{(\alpha,k)}=\sum_{\alpha',k'}\left(U_V^\dagger D_V^{1/2}\right)_{\alpha k,\alpha' k'}\tilde v_i^{(\alpha',k')}
\ee
and
\be
w_j^{(\alpha,k)}=\sum_{\alpha',k'}\left(U_W^\dagger D_W^{1/2}\right)_{\alpha k,\alpha' k'}\tilde w_j^{(\alpha',k')}.
\ee
Substitution into the expression (\ref{eq:newA}) for the new amplitudes yields
\be  \label{eq:AwithC}
\tilde A_{ij}=\sum_{\alpha',k'}\sum_{\alpha'',k''} C_{\alpha'k',\alpha''k''}
                            \tilde v_i^{(\alpha',k')}\tilde w_j^{(\alpha'',k'')},
\ee
with
\be
C_{\alpha'k',\alpha''k''}=\sum_{\alpha,k}\left(D_V^{1/2}U_V^*\right)_{\alpha'k',\alpha k}
                                  \left(U_W D_W^{1/2}\right)_{\alpha k,\alpha''k''}.
\ee
As a rank-two matrix, $C$ is just $D_V^{1/2}U_V^*U_W^\dagger D_W^{1/2}$, to which SVD
can be applied, to obtain $C=U_L^T\Lambda' U_R$.  Here $\Lambda'$ is diagonal with
entries $\lambda'_n$, again decreasing in magnitude.  For matrix elements of $C$, we then have
\be
C_{\alpha'k',\alpha''k''}=\sum_{n=1}^{\nsvd N_H}\left(U_L^T\right)_{\alpha'k',n}\lambda'_n
                                \left(U_R\right)_{n,\alpha''k''}.
\ee
The sum over $n$ is truncated to the first $\nsvd$ terms.  Substitution into (\ref{eq:AwithC})
then yields the compressed approximation to the amplitudes
\be
\tilde A_{ij}\simeq\sum_{n=1}^{\nsvd}\lambda'_n
       \sum_{\alpha,k}\left(U_L D_V^{-1/2}U_V\right)_{n,\alpha k}v_i^{(\alpha,k)}
       \sum_{\alpha',k'}\left(U_R D_W^{-1/2}U_W\right)_{n,\alpha' k'}w_j^{(\alpha',k')}.
\ee
This brings the action of the Hamiltonian back to the compressed SVD form with
$\nsvd$ terms.

\subsection{Diagonalization}
\label{sec:diagonalization}

For a matrix eigenvalue problem $H\vec{\psi}=\lambda\vec{\psi}$, where the matrix $H$
is not stored but instead computed as needed, a natural choice for the diagonalization
method is the Lanczos algorithm~\cite{Lanczos}.  In the case of a complex symmetric
matrix, such as will occur in the model discussed below, the algorithm generates
a sequence of vectors $\{\vec{u}_n\}$ from an initial guess $\vec{u}_1$ by
the following steps:
\bea
\vec{v}_{n+1}&=&H\vec{u}_n-b_n\vec{u}_{n-1}\;\;\;
(\mbox{with}\;b_1=0) \nonumber \\
a_n&=&\vec{v}_{n+1}\cdot\vec{u}_n \nonumber \\
\vec{v}_{n+1}^{\,\prime}&=&\vec{v}_{n+1}-a_n\vec{u}_n \\
b_{n+1}&=&\sqrt{\vec{v}_{n+1}^{\,\prime}\cdot\vec{v}_{n+1}^{\,\prime}}
\nonumber \\
\vec{u}_{n+1}&=&\vec{v}_{n+1}^{\,\prime}/b_{n+1}\,.   \nonumber
\eea
The dot products do not use conjugation, which leaves $a_n$ and $b_n$ possibly
complex. The dot product between two vectors stored in compressed SVD form,
such as
\be
A_{ij}=\sum_{k=1}^{\nsvd} \lambda_k V_{ik} W_{jk} \;\; \mbox{and} \;\;
B_{ij}=\sum_{k=1}^{\nsvd} \lambda'_k V'_{ik} W'_{jk},
\ee
can be written as
\be
\vec{A}\cdot\vec{B}=\sum_{k,k'}^{\nsvd} \lambda_k\lambda_{k'}
    \sum_i V_{ik}V'_{ik'} \sum_j W_{jk}W'_{jk'}.
\ee

These steps produce a complex symmetric, tridiagonal representation 
$T$ of the original matrix, with $a_n$ as the diagonal elements and $b_n$ 
the off-diagonal elements, with respect to the orthonormal vectors $\vec{u}_n$.
Diagonalization of $T$ then provides the eigenvalues of $H$; if the iterations
are stopped before a complete orthonormal basis is generated (which is usually
the case), the eigenvalues of $T$ approximate those of $H$, with the largest
and smallest approximated best.

The basis of Lanczos vectors $\vec{u}_n$ is orthonormal only if exact
arithmetic is used in the computations.  As is well known, round-off error
will gradually destroy the orthogonality and allow copies of previous Lanczos
vectors to enter the set, along with copies of eigenvalues in the diagonalization
of $T$~\cite{Lanczos}.

This de-orthogonalization is made much worse by SVD compression of the Lanczos
vectors.  Not only is each multiplication by $H$ approximated through compression,
as described in the previous subsection, but each subtraction of compressed 
vectors also requires an additional compression step, to keep the resulting
vector at the same level of compression.  The errors introduced by these
compressions rapidly destroy orthogonality and render the Lanczos process
useless.

Instead, we use a power method with forced orthogonalization, as do 
Weinstein {\em et al.}~\cite{Weinstein}.  From an initial guess $\vec{v}_1$,
a sequence of vectors $\vec{v}_n=H\vec{v}_{n-1}$ is generated
up to order $n=b$.  This set of $b$ vectors, all in compressed SVD form,
is orthogonalized by applying SVD to the overlap matrix 
$\langle \vec{v}_n|\vec{v}_{n'}\rangle$, to obtain an orthonormal set
$\{\vec{u}_1,\vec{u}_2,\ldots,\vec{u}_b\}$.  The matrix representation
of $H$ in this basis, $H_{ij}=\langle\vec{u}_i|H|\vec{u}_j\rangle$,
is diagonalized.  The smallest eigenvalue and its corresponding eigenvector
are extracted, and the eigenvector is used as a new $\vec{v}_1$ to
generate a new sequence.  The process is repeated until the eigenvalue
converges.  The value of $b$ is kept small, to minimize storage
requirements for the sequence of vectors; for results reported here, 
we used $b=3$.

\section{Sample application}
\label{sec:application}

\subsection{Model theory}
\label{sec:model}

As an illustration of the use of the compression and diagonalization algorithms,
we apply them to the DLCQ approximation of a simple model.  This model was
previously considered by Brodsky {\em et al.}~\cite{bhm} without compression.
They constructed the model as the light-front analog of the Greenberg--Schweber
static-source model~\cite{Greenberg}, designed to have an analytic solution.
It can be viewed as a heavy-fermion limit of the Yukawa model, where the
fermion has no true dynamics of its own and acts only as a source and
sink for bosons, without changing its spin.

The light-cone Hamiltonian for the model is
\bea
\lefteqn{\Pminus=
       \int\frac{dp^+d^2p_\perp}{16\pi^3p^+}(\frac{M_0^2}{P^+}+M'_0\frac{p^+}{P^+})
                 \sum_\sigma b_{\ub{p}\sigma}^\dagger
                                  b_{\ub{p}\sigma}} \hspace{0.5in} \\
  & +&\int\frac{dq^+d^2q_\perp}{16\pi^3q^+}
       \left[\frac{\mu^2+q_\perp^2}{q^+}
                       a_{\ub{q}}^\dagger a_{\ub{q}}
           + \frac{\mu_1^2+q_\perp^2}{q^+}
                        a_{1\ub{q}}^\dagger a_{1\ub{q}}
               \right]  \nonumber \\
    & +&\frac{g}{P^+}\int\frac{dp_1^+d^2p_{\perp1}}{\sqrt{16\pi^3p_1^+}}
            \int\frac{dp_2^+d^2p_{\perp2}}{\sqrt{16\pi^3p_2^+}}
              \int\frac{dq^+d^2q_\perp}{16\pi^3q^+}
                \sum_\sigma b_{\ub{p}_1\sigma}^\dagger
                             b_{\ub{p}_2\sigma}
   \nonumber \\
     & &\times \left[
      \left(\frac{p_1^+}{p_2^+}\right)^\gamma
         a_{\ub{q}}^\dagger
              \delta(\ub{p}_1-\ub{p}_2+\ub{q})
        +\left(\frac{p_2^+}{p_1^+}\right)^\gamma
        a_{\ub{q}}
              \delta(\ub{p}_1-\ub{p}_2-\ub{q}) \right.
    \nonumber \\
     & & \left.
       +i\left(\frac{p_1^+}{p_2^+}\right)^\gamma
       a_{1\ub{q}}^\dagger
             \delta(\ub{p}_1-\ub{p}_2+\ub{q})
      +i\left(\frac{p_2^+}{p_1^+}\right)^\gamma
      a_{1\ub{q}}
            \delta(\ub{p}_1-\ub{p}_2-\ub{q}) \right],
    \nonumber
\eea
where $b_{\ub{p}\sigma}^\dagger$ creates a fermion with momentum $\ub{p}$ 
and spin $\sigma$, $a_{\ub{q}}^\dagger$ creates a boson with momentum $\ub{q}$,
and $a_{1\ub{q}}^\dagger$ creates a heavy Pauli--Villars (PV) boson with
momentum $\ub{q}$.  The PV boson is included to regulate the model at
large transverse momenta~\cite{bhm,PauliVillars}.  The subtractions
necessary for the regularization are arranged by assigning an imaginary
coupling to the PV boson.  The bare mass of the fermion is $M_0^2$; the
term with $M'_0$ is a counterterm needed to subtract against the fermion
self-energy.  The masses of the physical and PV bosons are $\mu$ and $\mu_1$,
respectively.  With no fermion loops allowed, the boson masses are not 
renormalized.  The parameter $\gamma$ allows for a continuum of models;
however, we consider only the most natural value, $\gamma=1/2$, for
which the small-momentum behavior of the wave functions is uniformly 
a square-root dependence.

A mass eigenstate of this Hamiltonian, the one that represents the fermion
dressed by a cloud of bosons, can be expanded in Fock states as
\bea
|\Phi_\sigma(\ub{P})\rangle&=&\sqrt{16\pi^3P^+}\sum_{n,n_1}
                    \int\frac{dp^+d^2p_\perp}{\sqrt{16\pi^3p^+}}
   \prod_{i=1}^n\int\frac{dq_i^+d^2q_{\perp i}}{\sqrt{16\pi^3q_i^+}}
   \prod_{j=1}^{n_1}\int\frac{dr_j^+d^2r_{\perp j}}{\sqrt{16\pi^3r_j^+}} \\
   &  & \times \delta(\ub{P}-\ub{p}
                     -\sum_i^n\ub{q}_i-\sum_j^{n_1}\ub{r}_j)
       \phi^{(n,n_1)}(\ub{q}_i,\ub{r}_j;\ub{p})
         \frac{1}{\sqrt{n!n_1!}}b_{\ub{p}\sigma}^\dagger
          \prod_i^n a_{\ub{q}_i}^\dagger 
             \prod_j^{n_1} a_{1\ub{r}_j}^\dagger |0\rangle .
   \nonumber
\eea
Its normalization is
\be
\langle\Phi_\sigma(\ub{P}')|\Phi_\sigma(\ub{P}\rangle
=16\pi^3P^+\delta(\ub{P}'-\ub{P}),
\ee
which implies the following condition for the Fock-state wave functions:
\be  \label{eq:NormCondition}
1=\sum_{n,n_1}\prod_i^n\int\,dq_i^+d^2q_{\perp i}
                     \prod_j^{n_1}\int\,dr_j^+d^2r_{\perp j}
    \left|\phi^{(n,n_1)}(\ub{q}_i,\ub{r}_j;
           \ub{P}-\sum_i\ub{q}_i
                              -\sum_j\ub{r}_j)\right|^2.
\ee

We keep to a frame where the total transverse momentum $\vec{P}_\perp$
is zero and the mass eigenvalue problem is
\be
\Pminus|\Phi_\sigma(\ub{P})\rangle=\frac{M^2}{P^+}|\Phi_\sigma(\ub{P})\rangle.
\ee
This leads to a coupled set of integral equations for the wave functions~\cite{bhm}
\bea \label{eq:CoupledEqns}
\lefteqn{\left[M^2-M_0^2-M'_0p^+
  -\sum_i\frac{\mu^2+q_{\perp i}^2}{y_i}
                  -\sum_j\frac{\mu_1^2+r_{\perp j}^2}{z_j}\right]
                    \phi^{(n,n_1)}(\ub{q}_i,
                           \ub{r}_j,\ub{p})} \hspace{0.2in} \\
& =g&\left\{\sqrt{n+1}\int\frac{dq^+d^2q_\perp}{\sqrt{16\pi^3q^+}}
              \left(\frac{p^+-q^+}{p^+}\right)^\gamma
              \phi^{(n+1,n_1)}(\ub{q}_i,\ub{q},
                    \ub{r}_j,\ub{p}-\ub{q})\right.
\nonumber \\
& & +\frac{1}{\sqrt{n}}\sum_i\frac{1}{\sqrt{16\pi^3q_i^+}}
              \left(\frac{p^+}{p^++q_i^+}\right)^\gamma
              \phi^{(n-1,n_1)}(\ub{q}_1,\ldots,\ub{q}_{i-1},
                      \ub{q}_{i+1},\ldots,\ub{q}_n,
                       \ub{r}_j,\ub{p}+\ub{q}_i)
\nonumber \\
& &+i\sqrt{n_1+1}\int\frac{dr^+d^2r_\perp}{\sqrt{16\pi^3r^+}}
              \left(\frac{p^+-r^+}{r^+}\right)^\gamma
              \phi^{(n,n_1+1)}(\ub{q}_i,\ub{r}_j,
                           \ub{r},\ub{p}-\ub{r})
\nonumber \\
& & +\left.\frac{i}{\sqrt{n_1}}\sum_j\frac{1}{\sqrt{16\pi^3r_j^+}}
              \left(\frac{p^+}{p^++r_j^+}\right)^\gamma
              \phi^{(n,n_1-1)}(\ub{q}_i,\ub{r}_1,\ldots,
                                     \ub{r}_{j-1},
                        \ub{r}_{j+1},\ldots,\ub{r}_{n_1},
                           \ub{p}+\ub{r}_j) \right\},
\nonumber
\eea
with $y_i=q_i^+/P^+$ and $z_j=r_j^+/P^+$.  There is an analytic solution
\be   \label{eq:AnalyticSoln}
\phi^{(n,n_1)}=\sqrt{Z}\frac{(-g)^n(-ig)^{n_1}}{\sqrt{n!n_1!}}
         \left(\frac{p^+}{P^+}\right)^\gamma
         \prod_i\frac{y_i}{\sqrt{16\pi^3q_i^+}(\mu^2+q_{\perp i}^2)}
          \prod_j\frac{z_j}{\sqrt{16\pi^3r_j^+}(\mu_1^2+r_{\perp j}^2)},
\ee
with $M_0=M$ and
\be
M'_0=\frac{g^2/P^+}{16\pi^2}\frac{\ln\mu_1/\mu}{\gamma+1/2},
\ee
which guided the earlier work~\cite{bhm}.

The DLCQ approximation uses as a basis the discrete set of
plane waves with (anti)periodic boundary conditions on the 
light-cone box
\be
-L<x^-<L\,,\;\; -L_\perp<x,y<L_\perp.
\ee
The allowed momenta are then the discrete set
\be
p^+\rightarrow\frac{\pi}{L}n\,, \;\;
\vec{p}_\perp\rightarrow
     (\frac{\pi}{L_\perp}n_x,\frac{\pi}{L_\perp}n_y)\,,
\ee
with $n$ even for bosons, corresponding to periodic boundary
conditions, and $n$ odd for fermions and antiperiodic boundary
conditions.  The total longitudinal momentum $P^+$ for the dressed fermion
defines an odd integer $K$, called the resolution~\cite{PauliBrodsky},
such that $P^+=\pi K/L$.  The longitudinal momentum fractions are then
all of the form $n/K$, and $1/K$ sets the longitudinal resolution of
the approximation.  Integrals are approximated by the trapezoidal rule
\be \label{eq:rawDLCQ}
\int dp^+ \int d^2p_\perp f(p^+,\vec{p}_\perp)\simeq
   \frac{2\pi}{L}\left(\frac{\pi}{L_\perp}\right)^2
   \sum_n\sum_{n_x,n_y=-N_\perp}^{N_\perp}
   f(n\pi/L,\vec{n}_\perp\pi/L_\perp)\,.
\ee
This yields a matrix approximation for the coupled integral
equations of the model~\cite{bhm}
\bea   \label{eq:MatrixEq}
\lefteqn{\left[M^2-M_0^2-M'_0\frac{n}{K}
    -\sum_i\frac{\mu^2+\pi^2(m_{ix}^2+m_{iy}^2)/L_\perp^2}{m_i/K} \right. } && \\
 && \rule{1.5in}{0mm} \left.
  -\sum_j\frac{\mu_1^2+\pi^2(l_{jx}^2+l_{jy}^2)/L_\perp^2}{l_j/K}
       \right]\widetilde\psi^{(n,n_1)}(\ub{m}_i,\ub{l}_j,\ub{n})
       \nonumber \\
&  =&\frac{g\pi}{L_\perp\sqrt{8\pi^3}}
   \left\{\sum_{\ub{m}}
        \frac{1}{\sqrt{m}}
        \left(\frac{n-m}{n}\right)^\gamma
              \widetilde\psi^{(n+1,n_1)}(\ub{m}_i,\ub{m},
                    \ub{l}_j,\ub{n}-\ub{m})\right.
 \nonumber \\
& &  +\sum_i\frac{1}{\sqrt{m_i}}
        \left(\frac{n}{n+m_i}\right)^\gamma
              \widetilde\psi^{(n-1,n_1)}(\ub{m}_1,\ldots,\ub{m}_{i-1},
                        \ub{m}_{i+1},\ldots,\ub{m}_n,
                          \ub{l}_j,\ub{n}+\ub{m}_i)
\nonumber \\
& &+i\sum_{\ub{l}}
        \frac{1}{\sqrt{l}}
        \left(\frac{n-l}{n}\right)^\gamma
              \widetilde\psi^{(n,n_1+1)}(\ub{m}_i,\ub{l}_j,
                      \ub{l},\ub{n}-\ub{l})
\nonumber \\
& & +\left.i\sum_j\frac{1}{\sqrt{l_j}}
        \left(\frac{n}{n+l_j}\right)^\gamma
              \widetilde\psi^{(n,n_1-1)}(\ub{m}_i,\ub{l}_1,\ldots,
                      \ub{l}_{j-1},\ub{l}_{j+1},\ldots,
              \ub{l}_{n_1},\ub{n}+\ub{l}_j) \right\},
\nonumber
\eea
with $\ub{n}=\ub{K}-\sum_i\ub{m}_i-\sum_j\ub{l}_j$, $\ub{K}=(K,\vec{0}_\perp)$, and
\be
\widetilde\psi^{(n,n_1)}=\sqrt{n!n_1!}
             \left[\frac{2\pi}{L}
                \left(\frac{\pi}{L_\perp}\right)^2\right]^{(n+n_1)/2}\phi^{(n,n_1)}.
\ee
The normalization of the discrete amplitudes $\widetilde\psi^{(n,n_1)}$ is
\be
1=\sum_{n,n_1}  \sum_{\ub{m}_1}\,^\prime \cdots \sum_{\ub{m}_n}\,^\prime
   \sum_{\ub{l}_1}\,^\prime \cdots \sum_{\ub{l}_{n_1}}\,^\prime
   \frac{\left|\widetilde\psi^{(n,n_1)}\right|^2}
        {N_{\{\underline{m}_i\}}N_{\{\underline{l}_j\}}},
\ee
where the prime on the sum $\sum_{\underline{m}_i}^\prime$ restricts
the momenta $\ub{m}_1,\ldots,\ub{m}_n$ to one ordering and where the factorials
$N_{\{\underline{m}_i\}}\equiv
N_{\underline{m}_1}!N_{\underline{m}_2}!\cdots$, with
$N_{\underline{m}_1}$ the number of times that
$\underline{m}_1$ appears in the collection
$\{\underline{m}_i\}$, take into account multiplicities of 
identical particles.   We include these factors in the normalization
rather than in the coupled equations, so that the Hamiltonian matrix
can factorize between transverse and longitudinal momenta.

\subsection{Results}
\label{sec:results}

In order to compute results, we must first have a finite matrix
representation.
At fixed resolution $K$, the matrix representation is finite
in size with respect to the longitudinal momenta. This is 
because all longitudinal momenta are positive.  For a state
with bosons of longitudinal momentum indices $m_i$, PV bosons
with indices $l_j$ and fermion index $n$, $K$ must equal
$n+\sum_i m_i+\sum_j l_j$.  All indices are positive, and therefore
the range of the sums must be finite.  Consequently, at fixed $K$
the maximum number of bosons and PV bosons is finite.
Results are calculated for $K=11$ and 13, for which the
maximum number of bosons in a Fock state is five and six,
respectively.

For the transverse directions, we must impose a cutoff $N_\perp$
on the range of transverse indices.  The cutoff used in \cite{bhm}
was to limit the invariant mass of each particle: $(\mu^2+p_\perp^2)/p^+<\Lambda^2$.
We cannot use such a cutoff here because it does not factorize
between transverse and longitudinal momentum components.
Instead we directly limit the range $N_\perp$ of transverse
indices.  To keep the calculation small in size, we have
used only $N_\perp=1$ for physical and PV bosons, and $N_\perp=2$
for the fermion.  The total number of Fock states in the basis
is then 23,046 for $K=11$ and 89,739 for $K=13$.

Because we use a different cutoff, the numerical results can
only be compared with Ref.~\cite{bhm} in the limit of infinite
resolutions $K$ and $N_\perp$.  In this same limit we
should recover the analytic answer.  However, our purpose
is to evaluate the compression algorithm, not DLCQ itself.
Therefore, what we should check is whether there is convergence to
a numerical result at a particular set of resolutions
as the compression is removed, not whether the numerical result
converges to the exact answer.

The values used for the parameters of the model are taken
from \cite{bhm} at comparable resolutions.  These values are 
$M_0^2=0.8547\mu^2$ and $g=13.293\mu$ for $K=11$, and 
$M_0^2=0.8518\mu^2$ and $g=13.230\mu$ for $K=13$.
In both cases, $\mu_1^2=10\mu^2$ and $L_\perp=0.8165 \mu/\pi$.
The initial guess for the power method iterations was to set
the bare fermion amplitude to one and all other wave functions
to zero.

\begin{figure}[ht]
\vspace{0.2in}
\centerline{\includegraphics[width=14cm]{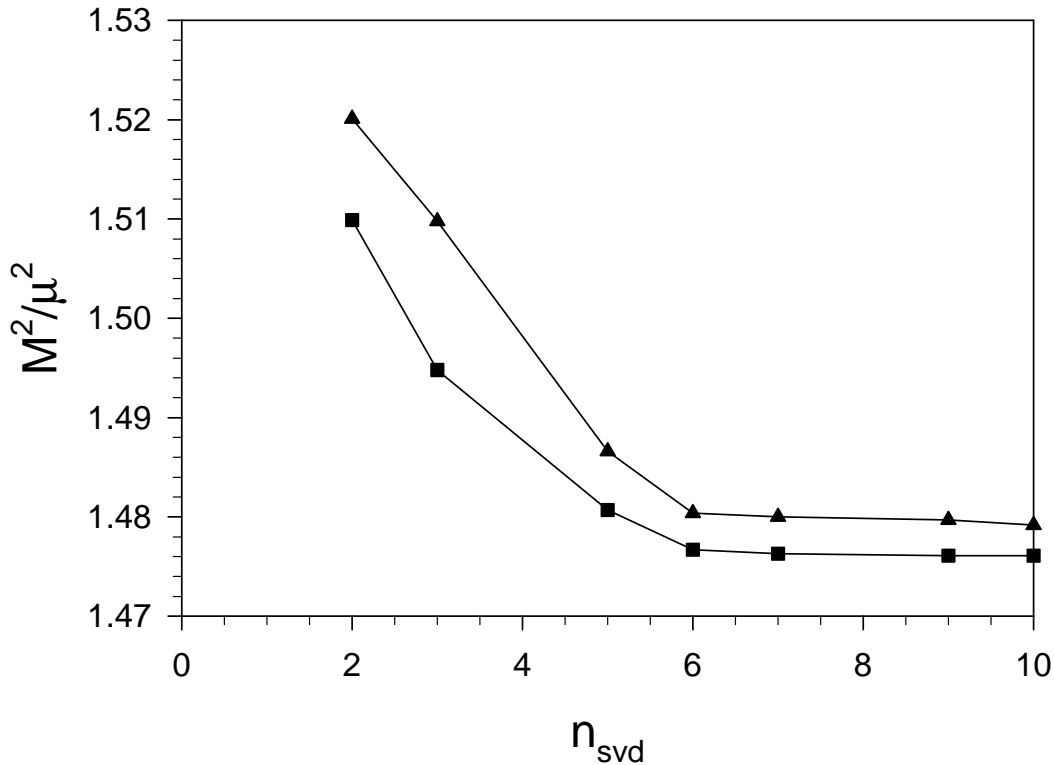}}
\caption{\label{fig:M2}
Mass eigenvalues $M^2$ as functions of the compression parameter $\nsvd$.
The DLCQ resolutions are $K=11$ for the filled triangles and
$K=13$ for the filled squares.  The masses are measured in units
of the constituent boson mass $\mu$.
}
\end{figure}

\begin{figure}[ht]
\vspace{0.2in}
\centerline{\includegraphics[width=14cm]{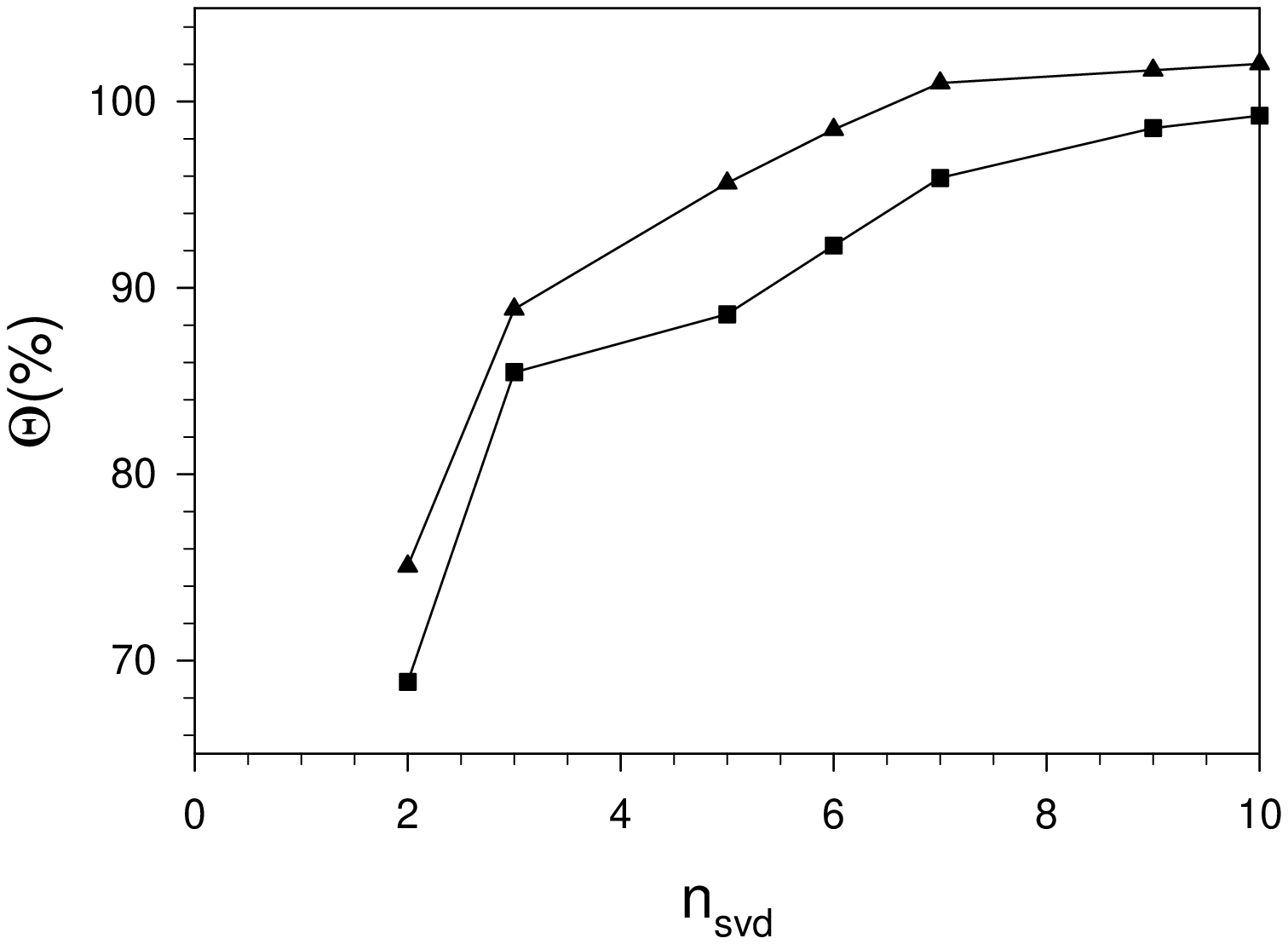}}
\caption{\label{fig:theta} 
Same as Fig.~\ref{fig:M2}, but for the compression ratio $\Theta$.
}
\end{figure}

The results for the smallest eigenvalue $M^2$, as a function of
the compression, are given in Fig.~\ref{fig:M2}.  As the compression
is removed, the eigenvalue clearly converges.  In Fig.~\ref{fig:theta}, 
we show the percentage reduction in memory achieved at each level
of compression.  This reduction is defined as the ratio of the
memory used for a compressed vector $M_{\nsvd}$ to that required
for a standard one-dimensional array $M_{1D}$:
\be
\Theta=\frac{M_{\nsvd}}{M_{1D}}\times 100\%
\ee
At relatively coarse compressions of $\sim 70\%$, the eigenvalue differs
from the converged value by $\sim3\%$.

In larger calculations, the compression ratio can be much smaller.
In these tests, the value of the longitudinal resolution $K$ is
not particularly high and limits the number of longitudinal states
in any one Fock sector to no more than 10 for $K=11$ and 15 for $K=13$.
The number of transverse states is much larger, at 2917 and 9093, respectively.
As $K$ is increased, the amount of compression can be increased.


\section{Summary}
\label{sec:summary}

We have converted the Weinstein compression algorithm~\cite{Weinstein}
to a form applicable to the DLCQ approximation~\cite{DLCQreview} of a quantum field theory.
As a test, we have applied it to a model theory~\cite{bhm} and investigated
the effects of compression on convergence of the mass eigenvalue and
on the memory requirements.  The results are shown in Figs.~\ref{fig:M2} and \ref{fig:theta}.
They indicate that in a larger calculation, of the sort required for 
good overall convergence, the memory savings could be substantial, at the
small cost of a few percent in the accuracy of the answer.  There is,
of course, a significant increase in computational overhead.  Also,
these reductions in memory for vectors are only meaningful if the
Hamiltonian matrix elements are computed as needed; when the Hamiltonian
is stored, its memory requirements dwarf those of the vectors.

The diagonalization method paired with the compression algorithm
must be chosen carefully.  Ordinary Lanczos iterations fail because
they require vector subtractions, which can be computed only
approximately for compressed vectors; the usual loss of orthogonality
that occurs in the Lanczos method is greatly exacerbated by these
errors.  We instead used a power method combined with explicit
orthogonalization, as suggested in \cite{Weinstein}.  This was
sufficient to extract the smallest eigenvalue.

This work could be extended in various ways.  For the model
considered here, additional tests could be done at
larger values of $K$, if the Hamiltonian matrix is computed as needed,
as it must be in a real application, rather than stored, which
was done here for convenience.  Theories in 2+1 dimensions, including
a lower-dimensional version of this model, might be the more natural
place for this algorithm, because having one less transverse dimension
would bring the wave function matrices closer to the square shape
that is optimal for compression.  A first application to a real,
(3+1)-dimensional theory would naturally be to Yukawa theory, where, unlike
the present model, the fermion would have its own dynamics and the
interactions would include spin.  This could be compared with the
uncompressed DLCQ approximation studied by 
Brodsky {\em et al.}~\cite{DLCQYukawa}.

\acknowledgments
This work was supported in part by the Department of Energy
through Contract No.\ DE-FG02-98ER41087.





\begin{thebibliography}{}

\bibitem{Weinstein} M. Weinstein, A. Auerbach, and V.R. Chandra,
Phys.\ Rev.\ E {\bf 84}, 056701 (2011).

\bibitem{NumRecipes}
W.H. Press, S.A. Teukolsky, W.T. Vetterling, and B.P. Flannery, 
{\em Numerical Recipes: The Art of Scientific Computing},
(Cambridge, New York, 2007).

\bibitem{Lanczos} \label{ref:Lanczos}
   C. Lanczos,
   J. Res.\ Nat.\ Bur.\ Stand.\ {\bf 45}, 255 (1950);
   J. Cullum and R.A. Willoughby,
   J. Comput.\ Phys.\ {\bf 44}, 329 (1981);
   {\em Lanczos Algorithms for Large Symmetric Eigenvalue Computations}
   (Birkhauser, Boston, 1985), Vol.\ I and II.
 
\bibitem{PauliBrodsky} H.-C. Pauli and S.J. Brodsky,
Phys.\ Rev.\ D {\bf 32}, 1993 (1985); {\bf 32}, 2001 (1985).
   
\bibitem{DLCQreview} For reviews of light-cone quantization, see
   M. Burkardt, Adv.\ Nucl.\ Phys.\ {\bf 23}, 1 (2002);
   S.J. Brodsky, H.-C. Pauli, and S.S. Pinsky, 
   Phys.\ Rep.\ {\bf 301}, 299 (1998).
   
\bibitem{Dirac} {P.A.M. Dirac, 
   Rev.\ Mod.\ Phys.\ {\bf 21}, 392 (1949).}

\bibitem{bhm} S.J. Brodsky, J.R. Hiller, and G. McCartor,
  Phys.\ Rev.\ D {\bf 58}, 025005 (1998).

\bibitem{Greenberg} O.W. Greenberg and S.S. Schweber,
Nuovo Cimento {\bf 8}, 378 (1958).

\bibitem{PauliVillars} W. Pauli and F. Villars,
   Rev.\ Mod.\ Phys.\ {\bf 21}, 434 (1949).
   
\bibitem{DLCQYukawa} S.J.~Brodsky, J.R.~Hiller and G.~McCartor,
  Phys.\ Rev.\ D {\bf 64}, 114023 (2001).
  

\end{thebibliography}
\end{document}